\documentstyle[aps]{revtex}
\begin{document}
\draft
\title{On rigidly rotating perfect fluid cylinders}
\author{B.V.Ivanov\thanks{%
E-mail: boyko@inrne.bas.bg}}
\address{Institute for Nuclear Research and Nuclear Energy,\\
Tzarigradsko Shausse 72, Sofia 1784, Bulgaria}
\maketitle

\begin{abstract}
The gravitational field of a rigidly rotating perfect fluid cylinder with $%
\gamma $-law equation of state is found analytically. The solution has two
parameters and is physically realistic for $1.41<\gamma \leq 2$. Closed
timelike curves always appear at large distances.
\end{abstract}


\section{Introduction}

The metric for a stationary cylindrically symmetric spacetime has the
Papapetrou form \cite{one} 
\begin{equation}
ds^2=-e^{2U}\left( dt+Ad\varphi \right) ^2+e^{-2U}\left[ e^{2K}\left(
e^{2H}dr^2+dz^2\right) +W^2d\varphi ^2\right] ,  \label{one}
\end{equation}
where the metric functions depend only on the radial coordinate $r$. The
function $H$ determines the gauge and can always be scaled to the isotropic
form $H=0$ by changing $r$. The static case follows when $A=0$. When the
source of the gravitational field is a perfect fluid, the energy-momentum
tensor reads 
\begin{equation}
T_{\alpha \beta }=\left( \mu +p\right) u_\alpha u_\beta +pg_{\alpha \beta },
\label{two}
\end{equation}
where $\mu $ is the energy density, $p$ is the pressure and $u^\alpha $ is
the four-velocity of the fluid. Rigid rotation is characterized by vanishing
shear. The field equations are less than the unknowns, which allows to
prescribe an equation of state $p=p\left( \mu \right) $.

The form (2) vanishes in vacuum, $p=\mu =0$. Dust is another particular case
with $p=0$. Vacuum solutions with a cosmological constant $\Lambda $ are
obtained when $\mu =-p=\Lambda /\kappa $. Here $\kappa $ is the Einstein
constant. When $p=-\Lambda /\kappa $, $\mu =\rho +\Lambda /\kappa $ we have
dust solutions with density $\rho $ on a cosmological background.

The first solutions with a source given by Eq. (2) were found by Lanczos in
1924 \cite{two}. He solved the case of dust and discussed the case of dust
plus a cosmological constant. Several years later the general vacuum
solution was found by Lewis \cite{three}. The first global solution,
containing rotating dust interior, was given by van Stockum \cite{four} who
rediscovered the dust solution of Lanczos. A rotating universe with
cosmological constant and constant $p=\mu =-\Lambda /\kappa $ was obtained
by G\"odel \cite{five}. A would be new five-parameter solution \cite{six}
was shown much later \cite{seven} to be equivalent to the two-parameter
Lanczos solution.

The field equations for perfect fluids with non-zero pressure in the axially
symmetric case were first written by Tr\"umper \cite{eight}. The cylindrical
problem was studied by Krasinski who obtained six types of new solutions,
all but one expressed in terms of hypergeometric functions \cite
{nine,ten,eleven,twelve}. The equations of state of these models are rather
complicated. Meanwhile, in the static case a solution with the linear
equation of state

\begin{equation}
p=\left( \gamma -1\right) \mu +p_1,  \label{three}
\end{equation}
where $\gamma =6/5$, was found and the general problem with $p_1=0$ was
simplified \cite{thirteen}. The case $\gamma =4/3$, $p_1=0$ was solved next 
\cite{fourteen}. The general solution with vanishing $p_1$ was soon derived 
\cite{fifteen} and rediscovered in a more convenient form some time later 
\cite{sixteen}. The connection between the two was clarified in Ref. \cite
{seventeen}. The static solution with cosmological term, which corresponds
to $\gamma =0$, $p_1=0$ was derived too \cite{eighteen}. General static
solutions with $p_1\neq 0$, however, are accessible even today only
numerically, including the simplest case of constant $\mu $ \cite{nineteen}.

The stationary metrics were studied in the past 10 years along two basic
lines. The general cosmological solution, which is implicitly contained in
the Krasinski solutions \cite{twenty}, was obtained explicitly \cite
{twentyone,twentytwo,twentythree}. This made it possible to build an
exterior for the G\"odel metric \cite{twentyfour}. On the other hand,
solutions with $p_1=0$ and $\gamma $ in the interval allowed by the energy
conditions $(1,2]$ were found by trial and error by Davidson \cite
{twentyfive,twentysix,twentyseven}. They confirm the results of a
qualitative numerical study, using a dynamical systems approach \cite
{twentyeight}. Davidson discovered by the same method a solution with finite
radius having $\gamma =2/3$ and $p_1\neq 0$ \cite{twentynine} which he
matched to the external Lewis solution \cite{thirty}. Finally, Sklavenites
studied two models with complicated equations of state, which reduce to the
Kramer and Evans solutions in the static limit \cite{one}.

In this paper we find analytically the gravitational field of a rigidly
rotating perfect fluid cylinder with equation of state given by Eq. (3) with 
$p_1=0$. Connections are made with most of the results found in the past.

In Sec.II the system of field equations from Ref. \cite{one} is simplified
and the regularity conditions are discussed. In Sec.III it is solved
analytically after a proper choice for $H$. The interval $1<\gamma \leq 2$
is divided into four regions. Two of them possess realistic algebraic
solutions. They are separated by the exceptional point $\gamma =3/2$ where
the solution becomes exponential.

In Sec.IV some particular solutions, mentioned in the Introduction, are
obtained from the general solution when $p$ and $\mu $ are constant. The
dust solutions in vacuum and in a cosmological background are rederived for
completeness.

The properties of the general solution are studied in Sec.V. The
acceleration, the vorticity vector and the angular velocity of rotation of
the perfect fluid are found. The curvature invariants are regular. Closed
timelike curves (CTC) appear in all cases. Sec.VI contains some conclusions.

\section{Field equations and regularity conditions}

We shall use the system in Ref. \cite{one} which is in comoving coordinates.
Thus, $u^t=e^{-U}$, the other components vanish. All dependence on $H$ may
be hidden in the operator $^{\prime }\equiv e^{-H}\frac d{dr}$, which
possesses many of the basic properties of a derivative. Then the field
equations read 
\begin{equation}
A^{\prime \prime }+4A^{\prime }U^{\prime }-\frac{A^{\prime }W^{\prime }}W=0,
\label{four}
\end{equation}
\begin{equation}
\frac{W^{\prime \prime }}W=16\pi pe^{2\left( K-U\right) },  \label{five}
\end{equation}
\begin{equation}
U^{\prime \prime }+\frac{e^{4U}A^{\prime 2}}{2W^2}+\frac{U^{\prime
}W^{\prime }}W=4\pi \left( \mu +3p\right) e^{2\left( K-U\right) },
\label{six}
\end{equation}
\begin{equation}
-\frac{W^{\prime \prime }}W+2K^{\prime }\frac{W^{\prime }}W-2U^{\prime 2}+%
\frac{e^{4U}A^{\prime 2}}{2W^2}=0,  \label{seven}
\end{equation}
\begin{equation}
K^{\prime \prime }+U^{\prime 2}+\frac{e^{4U}A^{\prime 2}}{4W^2}=8\pi
pe^{2\left( K-U\right) },  \label{eight}
\end{equation}
\begin{equation}
p^{\prime }+\left( \mu +p\right) U^{\prime }=0.  \label{nine}
\end{equation}
We use units where $G=c=1$. Then $\kappa =8\pi $. The last equation follows
from the others. Together with the equation of state (3) this is a system of
6 independent differential equations for 7 unknowns $\mu ,p,U,K,A,W,H$. The
solution should depend on the arbitrary gauge function $H$. A regular
solution must be elementary flat. This means 
\begin{equation}
\lim\limits_{r\rightarrow 0}r^{-1}e^{U-K-H}\left(
e^{-2U}W^2-e^{2U}A^2\right) ^{1/2}=1.  \label{ten}
\end{equation}

Let us simplify the above system. Eq.(4) is easily integrated once 
\begin{equation}
A^{\prime }=a_0We^{-4U},  \label{eleven}
\end{equation}
where $a_i$ are constants. This holds for any equation of state and
simplifies the terms with $A$ in the other equations. When $\gamma \neq 1$
Eqs. (3) and (9) give 
\begin{equation}
p=p_0e^{-\frac \gamma {\gamma -1}U},\quad \mu =\frac{p_0}{\gamma -1}e^{-%
\frac \gamma {\gamma -1}U}.  \label{twelve}
\end{equation}
For stability reasons the constant $p_0>0$. It measures the central
pressure. Eq. (6) can be integrated next 
\begin{equation}
WU^{\prime }-\frac{3\gamma -2}{4\left( \gamma -1\right) }W^{\prime }=-\frac{%
a_0}2A+a_1.  \label{thirteen}
\end{equation}
Eqs. (7,8) give an equation for $K$%
\begin{equation}
WK^{\prime }-W^{\prime }=-\frac{a_0}2A+a_2.  \label{fourteen}
\end{equation}
Eqs. (13,14) show that $K-U$ may be expressed through $W$. We have reduced
the initial system to Eqs. (5,11-14), all but one being of first order.

In a realistic solution $U,K$ and $H$ should approach constants when $%
r\rightarrow 0$. Then Eq. (10) shows that $W\sim r^\nu ,A\sim r^\tau $.
However, Eq. (11) gives $\tau =\nu +1$. Therefore, the term with $A$
decouples in this limit and we get the static case condition 
\begin{equation}
\lim_{r\rightarrow 0}e^{-K-H}\frac{\left| W\right| }r=1.  \label{fifteen}
\end{equation}
Hence, $W\sim r$ and $A$ may be written as 
\begin{equation}
A=\frac{qr^2}{1+F\left( r\right) },  \label{sixteen}
\end{equation}
where $q$ is a constant and $F\left( 0\right) =0$.

\section{The general solution}

The key to the solution of the simplified system of equations is the choice
of gauge. Different gauges have been proposed in the literature but none of
them works well here. Let us choose the following $H$%
\begin{equation}
e^H=\frac{a_0hW_r}{16\pi p_0\left( A+h_0\right) },  \label{seventeen}
\end{equation}
where $h,h_0$ are constants. The introduction of $h_0$ is necessary to
ensure the regularity of $e^H$ on the axis of rotation. We also require $%
H\left( 0\right) =0$. It is possible to satisfy Eq.(17) by fixing
appropriately most of the constants without fixing $H$. Therefore the
solution will depend on an arbitrary function.

Eq.(5) becomes a relation between $K$ and $U$%
\begin{equation}
he^{2K}=e^{\frac{2-\gamma }{\gamma -1}U},  \label{eighteen}
\end{equation}
which shows that $h>0$. Then Eqs. (13,14) should be checked for
compatibility. This is achieved by adjusting $a_2$ and $h$%
\begin{equation}
a_2=\frac{2-\gamma }{2\left( \gamma -1\right) }a_1+\frac{4-3\gamma }{4\left(
\gamma -1\right) }a_0h_0,  \label{nineteen}
\end{equation}
\begin{equation}
h=\frac{8\pi p_0}{a_0^2}f\left( \gamma \right) ,\qquad f\left( \gamma
\right) =\frac{11\gamma ^2-24\gamma +12}{\left( \gamma -1\right) \left(
3\gamma -4\right) }.  \label{twenty}
\end{equation}
The zeroes of the nominator of $f$ are $\gamma _{1,2}=\frac{12\pm 2\sqrt{3}}{%
11}$. The second is outside the realistic range, while the first is 
\begin{equation}
\frac{12+2\sqrt{3}}{11}=\frac{2\sqrt{3}}{2\sqrt{3}-1}=1.4065.
\label{twentyone}
\end{equation}
The sign of $f$ divides the range of $\gamma $ into three regions: $\left(
1,4/3\right) $ and $\left( 1.41,2\right) $ where $f>0$ and $\left(
4/3,1.41\right) $ where $f<0$. The sign of $p_0$ must coincide with the sign
of $f$, therefore in the third region the pressure is negative and
unrealistic. Also $f$ blows up at $\gamma =4/3$. We can arrange for $U$ and $%
H$ to vanish on the axis. Eq.(18) then shows that the same is possible for $K
$ only when $h=1$. This condition is a relation between the two independent
parameters $a_0$ and $p_0$ of the solution 
\begin{equation}
p_0f=\frac{a_0^2}{8\pi }.  \label{twentytwo}
\end{equation}
It also means that $K\sim U$. These restrictions characterize the Davidson
solutions.

In general, $e^K\rightarrow h^{-1/2}$ and $W\rightarrow h^{-1/2}r$ (from Eq.
(15)) on the axis. Then Eq. (17) fixes $h_0$ while Eq. (11) yields $q$%
\begin{equation}
h_0=\frac 18\sqrt{\frac{2f}{\pi \left| p_0\right| }}\frac{signa_0}{signp_0}%
,\qquad q=\frac{a_0}{2\sqrt{h}}.  \label{twentythree}
\end{equation}

Let us introduce 
\begin{equation}
Y\equiv \left( A+h_0\right) ^2.  \label{twentyfour}
\end{equation}
Inserting Eq. (17) into Eq. (11) and taking $W$ as an independent variable,
we get 
\begin{equation}
U_W=-\frac{Y_{WW}}{4Y_W}+\frac 1{4W}.  \label{twentyfive}
\end{equation}
The last equation to be satisfied is (13). Taking into account Eqs. (17,25)
it becomes a second-order non-linear ordinary differential equation for $%
Y\left( W\right) $: 
\begin{equation}
Y_{WW}=\left( b_0+\frac{c_0}{Y^{1/2}}\right) \frac{Y_W}W,  \label{twentysix}
\end{equation}
\begin{equation}
b_0=\frac{5\gamma -8}{3\gamma -4},\qquad c_0=-f\left( \frac{2a_1}{a_0}%
+h_0\right) .  \label{twentyseven}
\end{equation}
The parameter $c_0$ determines $a_1$. Eq. (26) is a generalized Emden-Fowler
equation \cite{thirtyone}. It belongs to the class 2.6.2.98 and,
fortunately, is among the 147 integrable cases. The parametric solution
hints at the possibility to reverse the variables. Writing $W=c_1e^G$, where 
$c_1$ is another constant, one obtains an integrable equation for $G$ which
yields 
\begin{equation}
G=\int \frac{dY}{c_2+\left( b_0+1\right) Y+2c_0Y^{1/2}}  \label{twentyeight}
\end{equation}
with yet another free constant $c_2$. This determines $W$ as a function of $A
$. Eq. (17) becomes 
\begin{equation}
e^H=\left( 2a_0\right) ^{-1}\frac{W_AA_r}{A+h_0}.  \label{twentynine}
\end{equation}
The parameter $a_0$ measures the effects of the fluid's rotation through Eq.
(16) 
\begin{equation}
A=\frac{\left( a_0r\right) ^2}{2\sqrt{8\pi p_0f}\left( 1+F\right) }.
\label{thirty}
\end{equation}

The static case is given by $a_0=0$. Obviously Eq. (17) or (29) is tied to
the stationary case. In principle, $A$ may be expressed through $H$ but an
integration is necessary, which depends on $F$. It is much easier to give
all fluid characteristics as functions of $A$. Next we determine $U$ from
Eqs. (11),(29) 
\begin{equation}
e^{4U}=\frac{fWW_A}{2\left( A+h_0\right) }.  \label{thirtyone}
\end{equation}
The function $K$ follows from Eq. (18) and is linear in $U$. The free
parameters for each $\gamma $ are two: $a_0$ (measures rotation) and $p_0$
(measures the central pressure and density $\mu _0=p_0/\left( \gamma
-1\right) $). The Davidson solutions have just one parameter because $p_0$
and $a_0$ are related by Eq.(22) like the mass and the angular momentum of
the Kerr solution.

The integral in Eq. (28) has two cases: $b_0+1=0$ ($\gamma =3/2$) and $%
b_0+1\neq 0$ ($\gamma \neq 3/2$). In the second case it is given by two
expressions, depending on the sign of $c_0^2-\left( b_0+1\right) c_2$. The
condition $W\sim r$ near the axis is very restrictive and tedious
calculations show that $W$ looses its dependence on $c_0$ and $c_2$ 
\begin{equation}
W=W_0A^{1/2}B^{\frac{\gamma -1}{2\left( 2\gamma -3\right) }},\qquad \left|
W_0\right| =\frac{\sqrt{2}}{h^{1/4}\left| a_0\right| ^{1/2}},
\label{thirtytwo}
\end{equation}
\begin{equation}
B=1+aA,\qquad a=\frac{\left( 2\gamma -3\right) }{\left( 3\gamma -4\right) h_0%
}.  \label{thirtythree}
\end{equation}
We can always choose $a_0>0,q>0$ and $W_0>0$ so that $A\geq 0,B>0$. The
metric functions and the pressure read 
\begin{equation}
e^H=\frac{A_r}{2\sqrt{qA}}B^{\frac{5-3\gamma }{2\left( 2\gamma -3\right) }%
},\qquad e^{2U}=B^{\frac{2-\gamma }{2\left( 2\gamma -3\right) }},
\label{thirtyfour}
\end{equation}
\begin{equation}
p=p_0B^{-\frac{\gamma \left( 2-\gamma \right) }{4\left( \gamma -1\right)
\left( 2\gamma -3\right) }},\qquad e^{2K}=h^{-1}B^{\frac{\left( 2-\gamma
\right) ^2}{4\left( \gamma -1\right) \left( 2\gamma -3\right) }}.
\label{thirtyfive}
\end{equation}
It is useful to write the metric in the so-called Lewis form, used in most
of the research on the topic 
\begin{eqnarray}
ds^2 &=&-B^{\frac{2-\gamma }{2\left( 2\gamma -3\right) }}dt^2-2AB^{\frac{%
2-\gamma }{2\left( 2\gamma -3\right) }}dtd\varphi +\left( W_0^2B-A\right)
AB^{\frac{2-\gamma }{2\left( 2\gamma -3\right) }}d\varphi ^2+\frac{A_r^2}{%
4hqA}B^{-\frac{9\gamma ^2-22\gamma +12}{4\left( 2\gamma -3\right) \left(
\gamma -1\right) }}dr^2+  \nonumber  \label{thirtysix} \\
&&\ +h^{-1}B^{-\frac{\left( 2-\gamma \right) \left( 3\gamma -4\right) }{%
4\left( 2\gamma -3\right) \left( \gamma -1\right) }}dz^2.  \label{thirtysix}
\end{eqnarray}

In the process of solution the gauge function was shifted effectively from $H
$ to $A$ and then to $F$. The simplest function should be chosen. However,
it is not possible to do this simultaneously in the whole region $1<\gamma
\leq 2$. It is divided into four parts.

When $3/2<\gamma \leq 2$ we can choose $F=0$. Now $f>0,a>0,p_0>0,$ hence $%
B\rightarrow \infty $, $p\rightarrow 0$ at infinity and that is physically
realistic. Setting $h=1$ we obtain the solution from Ref. \cite{twentysix}
or case a) from Ref. \cite{twentyseven}.

When $1.41<\gamma <3/2$, $f>0,p_0>0$ but $a<0$ and if $F=0$, $B$ will vanish
at some point, resulting in a singular metric. Let us take $F=-ar^2$, then $%
B=\left( 1-ar^2\right) ^{-1}$ is always positive and decreases monotonically
as $r\rightarrow \infty $. Then $p\rightarrow 0$ as necessary. The subcase $%
h=1$ leads to case c) from Ref. \cite{twentyseven}.

In the region $4/3<\gamma <1.41$, $f<0$ and to ensure the positivity of $h$
the central pressure should be negative. Then Eqs. (17), (30) make sense.
The appearance of tension, however, leads to instability with respect to
perturbations. When $h=1$ this case was discussed in \cite{twentyseven}. It
seems that no physical solution exists for this interval.

Finally, in the region $1<\gamma <4/3$, $f,p_0,a$ and $A/h_0$ are positive.
Then $B\geq 1$ according to Eq. (33). $B$ should decrease monotonically with
the distance in order that $p$ also decreases, but this is not possible
since $B\left( 0\right) =1$. The solution can not have a regular axis, in
accord with the results of Refs.\cite{twentyseven,twentyeight}. One can find
a solution whose other features are regular by setting $A\sim \left(
1+k^2r^2\right) ^{-1}$ as was done in case d) from Ref. \cite{twentyseven}.

The general conclusion is that a regular solution exists only in the
relativistic interval $1.41<\gamma \leq 2$ and two different choices of
gauge are necessary to cover it on both sides of the exceptional point $%
\gamma =3/2$. The Davidson one-parameter metrics are obtained when one puts
the correct $F$ and $h=1$ into Eq. (36).

In the special case $\gamma =3/2$ the integral in Eq. (28) is simpler and
elementary flatness requires 
\begin{equation}
W=b_1A^{1/2}e^{\alpha A},\qquad \alpha =\sqrt{\frac{8\pi p_0}3},\qquad b_1=%
\frac{a_0}{\alpha \sqrt{q}},  \label{thirtyseven}
\end{equation}
where $b_1$ is determined, like $W_0$ before, by the condition $H\left(
0\right) =0$. Then 
\begin{equation}
e^H=\frac{A_r}{2\sqrt{qA}}e^{\alpha A}.  \label{thirtyeight}
\end{equation}
Eq (31) determines $U$ and $p,\mu ,K$ follow from it 
\begin{equation}
e^{2U}=\frac 12\sqrt{3\alpha }b_1e^{\alpha A},  \label{thirtynine}
\end{equation}
\begin{equation}
e^{2K}=\frac{a_0^2}{24\pi p_0}e^U,\qquad p=p_0e^{-3U},\qquad \mu =2p.
\label{forty}
\end{equation}
The basic difference from the general case is the change $B\rightarrow
e^{\alpha A}$. The simplest choice for $A=qr^2$ leads to a regular solution
with decreasing pressure towards infinity. It also possesses two parameters.
Davidson proposed a single parameter solution with $A=qr$ and $h=1$ \cite
{twentyseven}, but its gauge function blows up at the axis and it is not
elementary flat. One obtains $2$ on the R.H.S. of Eq. (15).

It should be stressed that the static subcase is not a limiting case of the
general solution. Eqs. (4,11) do not hold and Eq. (17) is meaningless. One
should start with the system (5-9) with $A=0$ and solve it a new. Bronnikov 
\cite{fifteen} used the gauge $e^H=W$, while Kramer \cite{sixteen} expressed
the metric components as functions of $U$, which can be derived from
Bronnikov's solution by changing the gauge \cite{seventeen}.

\section{Some particular solutions}

The previous formulas are inapplicable when $\gamma =1$ and the pressure
vanishes. Then Eq. (5) shows that one can choose $W=r$. No exotic gauge is
needed, so that $H=0$. Eq. (9) becomes $\mu U^{\prime }=0$ and comprises two
cases. When $\mu =0$ we have vacuum and the system of Eqs. (4-8,11) gives 
\begin{equation}
A^{\prime }=a_0re^{-4U},\,\qquad \frac{2K^{\prime }}r=2U^{\prime 2}-\frac{%
a_0^2}2e^{-4U},  \label{fortyone}
\end{equation}
\begin{equation}
U^{\prime \prime }+\frac{U^{\prime }}r+\frac{a_0^2}2e^{-4U}=0.
\label{fortytwo}
\end{equation}
Eq. (42) coincides with Eq. (A4) from Ref. \cite{thirtytwo}, one of the few
papers where the Papapetrou form of the metric is used.

In the dust case $\mu \neq 0$ and consequently $e^U=1$. Then 
\begin{equation}
A=\frac{a_0}2r^2,\qquad K=-\frac{a_0^2}8r^2,\qquad \mu =\frac{a_0^2}{8\pi }%
e^{\frac{a_0^2r^2}4},  \label{fortythree}
\end{equation}
which is the Lanczos solution \cite{two,four}.

Let us discuss next solutions with constant $p\neq 0$ and $\mu $. If Eq. (3)
holds then Eq. (35) holds too and $p,\mu $ are constant when $\gamma =0$ or $%
\gamma =2$, i.e. $p=\pm \mu $. On the contrary, when $p,\mu $ are constant $%
\left( \mu +p\right) U^{\prime }=0$ so that either $p=-\mu $ (cosmological
solutions) or $U=0$ and Eq.(6) shows that $K=const=0$. Then Eq.(11)
simplifies. The comparison of Eqs. (5,7) yields $a_0^2=32\pi p$. When
inserted in Eq. (6) this gives $p=\mu $. This is the G\"odel solution. Let
us derive these two solutions from the general formulas (32-36). The
cosmological constant is $\Lambda =-8\pi p_0$ in all cases that follow.

a) G\"odel solution. We have $\gamma =2$, $p=p_0=\mu $, $f=4$, $U=K=0$ and 
\begin{equation}
h=\frac{32\pi p_0}{a_0^2},\qquad h_0=\frac 1{\sqrt{8\pi p_0}},\qquad q=\frac{%
a_0^2}{8\sqrt{2\pi p_0}},  \label{fortyfour}
\end{equation}
\begin{equation}
B=1+aA,\qquad a=\sqrt{2\pi p_0},  \label{fortyfive}
\end{equation}
\begin{equation}
W=W_0\sqrt{AB},\qquad W_0^2=h_0.  \label{fortysix}
\end{equation}
The gauge is again $H=0$. This implies an equation for $A$: 
\begin{equation}
A_r=2\sqrt{qAB}.  \label{fortyseven}
\end{equation}
Its solution reads 
\begin{equation}
A^{1/2}=\frac 12\left( e^{lr+c}-a^{-1}e^{-lr-c}\right) ,  \label{fortyeight}
\end{equation}
where $l=a_0/2\sqrt{2}$, while $c$ is an integration constant. It is fixed
by the condition $A\left( 0\right) =0$ to $e^c=1/\sqrt{a}$ and the solution
has two parameters $a_0$ and $p_0$. It is an elementary flat subcase of the
Wright solution \cite{six}, Eq.(31). Then hyperbolic functions are obtained.
If we fix the two parameters in terms of one constant $b$ 
\begin{equation}
a_0=\frac{\sqrt{2}}b,\qquad p_0=\frac 1{16\pi b^2},  \label{fortynine}
\end{equation}
the one-parameter G\"odel solution follows \cite{twentyfour} 
\begin{equation}
ds^2=-dt^2-4\sqrt{2}b\sinh ^2\frac r{2b}dtd\varphi +4b^2\left( \sinh ^2\frac 
r{2b}-\sinh ^4\frac r{2b}\right) d\varphi ^2+dr^2+dz^2.  \label{fifty}
\end{equation}
.

b) Cosmological solution. We have $\gamma =0$, $p=p_0=-\mu $, $f=3$ and 
\begin{equation}
h=\frac{24\pi p_0}{a_0^2},\qquad h_0=\sqrt{\frac 3{32\pi p_0}},
\label{fiftyone}
\end{equation}
\begin{equation}
B=1+aA,\qquad a=\sqrt{6\pi p_0},\qquad l=\sqrt{aqh}=a,  \label{fiftytwo}
\end{equation}
\begin{equation}
W=W_0A^{1/2}B^{1/6},\qquad W_0^2=\frac 1{\sqrt{6\pi p_0}},
\label{fiftythree}
\end{equation}
\begin{equation}
e^{2U}=B^{-1/3},\qquad e^{2K}=h^{-1}B^{1/3}.  \label{fiftyfour}
\end{equation}
Let us fix the gauge demanding $H=U-K$. The equation for $A$ is very similar
to Eq. (47) 
\begin{equation}
A_r=2\sqrt{qhAB}.  \label{fiftyfive}
\end{equation}
Its solution is given by Eq. (48) with $a,l$ taken from Eq. (52), $c$ is
fixed as above. The solution has two parameters but $a_0$ enters only in $%
e^{2K}$ . Comparison can be made with Refs. \cite
{twentyone,twentytwo,twentythree} after the calculation of the quantity $%
\Psi =We^{K-U}$%
\begin{equation}
\Psi =\frac{W_0A_r}{2\sqrt{q}h}=C_1\cosh lr+C_2\sinh lr.  \label{fiftysix}
\end{equation}
Here $C_i$ are combinations of $p_0,a_0$. The second relation coincides with
Eq. (38) from Ref. \cite{twentyone}. The case of $\Lambda $ with opposite
sign is obtained by taking negative $p_0$. This makes $l$ purely imaginary
and the hyperbolic functions become trigonometric.

Finally, let us discuss dust solutions on a cosmological background. The
pressure is constant while $\mu $ is determined from the equations. We set
again $H=0$. Eq. (9) gives $U=0$ and then Eq. (11) becomes 
\begin{equation}
A^{\prime }=a_0W.  \label{fiftyseven}
\end{equation}
Replaced into Eq. (6) it turns into an expression for $\mu $ as a function
of $K$%
\begin{equation}
\mu =-3p_0+\frac{a_0^2}{8\pi }e^{-2K}.  \label{fiftyeight}
\end{equation}
Eq. (8) becomes an equation for $K$ which may be integrated once 
\begin{equation}
K^{\prime 2}=8\pi p_0e^{2K}-\frac{a_0^2}2K+2d_1.  \label{fiftynine}
\end{equation}
Here $d_1$ is an integration constant. The term with $p_0$ may be expressed
through $K$ and $W$ in another way from Eqs. (5,7). The comparison with the
last equation yields 
\begin{equation}
W=d_2K^{\prime },\qquad A=a_0d_2K+d_3.  \label{sixty}
\end{equation}
Everything is expressed through $K$ that satisfies the non-linear first
order differential equation (59). It was first found by Lanczos and can not
be solved analytically. However, we can change the variable from $r$ to $y$
where $K=-y^2$. Then 
\begin{equation}
e^{2K}dr^2=\frac{4e^{-2y^2}y^2dy^2}{K^{\prime 2}}  \label{sixtyone}
\end{equation}
and all metric components become functions of $y$ \cite{six}. Elementary
flatness fixes three of the constants 
\begin{equation}
d_1=-4\pi p_0,\qquad d_2=\frac{\sqrt{2}}{a_0},\qquad d_3=0  \label{sixtytwo}
\end{equation}
so that the solution has two parameters $p_{0,}a_0$ \cite{seven}, like the
general solution and the other particular solutions discussed in this
section.

This problem was discussed again recently \cite{thirtythree} but the system
of equations was not transcribed correctly from Ref. \cite{thirtyfour} and
the wrong conclusion that an explicit solution is impossible was stated. The
system in Iftime's Eq. (17) is completely different from Eqs. (4-9) or Eqs.
(57-60), the main error being in the analog of Eq. (57). The system (4-9) is
equivalent to that of Tr\"umper \cite{eight} and its complex reformulation 
\cite{thirtyfour}. We have checked this system also by GRTensor.

In conclusion, we have shown how many of the known particular solutions
follow from the general one or directly from the system (4-9). Fluids with $%
p_1\neq 0$ in Eq.(3) have finite radius. The main system may be simplified
again, but we have been unable to solve the master differential equation.

The properties of these special metrics have been studied extensively
through the years. Therefore, in the next section only the properties of the
general solution will be discussed.

\section{Properties of the general solution}

We have calculated with the help of GRTensor the characteristics of $%
u^\alpha $ in the metric (1). The expansion and shear vanish as it should be
for a rigid rotation. The acceleration has only a radial component $v^r$ and
its magnitude is 
\begin{equation}
v=\left( e^U\right) _re^{-K-H}.  \label{sixtythree}
\end{equation}
The vorticity vector has only a longitudinal component $\omega ^z$ and a
magnitude 
\begin{equation}
\omega =\frac{A_r}{2W}e^{3U-K-H}.  \label{sixtyfour}
\end{equation}
The angular velocity $\Omega =\omega \left( 0\right) $ of the rigid-body
rotation of the fluid is 
\begin{equation}
\Omega =qh=\sqrt{2\pi p_0f},  \label{sixtyfive}
\end{equation}
where Eqs. (15-18), (20) were used. Amazingly, $a_0$, which determines the
magnitude of $A$ and the degree of stationarity, does not enter this
formula. The angular velocity is entirely determined by the central
pressure. This issue is obscured in the Davidson's solutions because of the
relation (22). Of course, $a_0$ has its influence upon the radial
distribution of $v$ and $\omega $.

For the general solution these characteristics are 
\begin{equation}
v=\left| \frac{2-\gamma }{3\gamma -4}\right| \sqrt{\frac{2\pi pA}{h_0B}}%
,\qquad \omega =\sqrt{2\pi pf}.  \label{sixtysix}
\end{equation}
When $h=1$ and Eq. (35) is taken into account we obtain the expressions of
Davidson, his Eqs. (5.7) and (5.9) from \cite{twentysix} or Eqs. (2.16),
(2.17) from \cite{twentyseven}. The vector $v^\alpha $ vanishes both at the
axis and at infinity when $F$ is chosen accordingly and $1.41<\gamma \leq 2$%
. At the axis $\omega $ equals $\Omega $ while at infinity it vanishes
together with the pressure. The spin of the fluid vanishes at $\gamma =1.41$
because this is a zero of $f\left( \gamma \right) $.

In the special case $\gamma =3/2$ Eq. (65) still holds while 
\begin{equation}
v=k_1A^{1/2}e^{-\frac 34\alpha A},\qquad \omega =k_2e^{-\frac 34\alpha A},
\label{sixtyseven}
\end{equation}
where $k_i$ are constants depending on $a_0,p_0$. The behaviour of $v$ and $%
\omega $ is as in the general case.

We have studied also the behaviour of the independent curvature invariants.
They are regular at the axis. For a typical quadratic invariant the most
dangerous term at infinity has the form 
\begin{equation}
I=A^2B^{2w},\qquad w=\frac{-7\gamma ^2+18\gamma -12}{8\left( \gamma
-1\right) \left( 2\gamma -3\right) }.  \label{sixtyeight}
\end{equation}
Fortunately, $I\sim v^2$ and vanishes at infinity together with $v$. The
linear invariant $R$ has terms $I\sim v$ and so on. The Petrov type of the
general solution is I except for $\gamma =2$ where it is D \cite
{twentysix,twentyseven}.

Finally, let us discuss the appearance of CTC. This happens when $g_{\varphi
\varphi }$ becomes negative. From Eq. (1) and the expression for the general
solution it follows that 
\begin{equation}
g_{\varphi \varphi }=e^{-2U}\left( W^2-e^{4U}A^2\right) =AB^{\frac{3\gamma -4%
}{2\left( 2\gamma -3\right) }}\left( W_0^2-\frac AB\right) .
\label{sixtynine}
\end{equation}
The term in the last bracket is always negative for large distances. When $%
1.5<\gamma \leq 2$ we set $F=0$ and $g_{\varphi \varphi }<0$ if 
\begin{equation}
r^2>\frac{4\left( \gamma -1\right) f}{\gamma a_0^2}.  \label{seventy}
\end{equation}
When $1.41<\gamma <1.5$ we set $F=-ar^2$ and $g_{\varphi \varphi }<0$ if 
\begin{equation}
r>\frac 2{a_0}.  \label{seventyone}
\end{equation}
When $\gamma =1.5$ the second equality in Eq. (69) does not hold. We have 
\begin{equation}
W-e^{2U}A=b_1A^{1/2}e^{\alpha A}\left( 1-\frac{\sqrt{3\alpha }}2%
A^{1/2}\right) ,  \label{seventytwo}
\end{equation}
which turns negative if 
\begin{equation}
A>\frac 2{\sqrt{6\pi p_0}},\qquad r>\frac{2\sqrt{2}}{a_0}.
\label{seventythree}
\end{equation}
Thus, CTC appear in all cases and the bigger $a_0$, the closer to the axis
they stand.

\section{Conclusions}

In this paper we have found analytically the gravitational field of a
rigidly rotating perfect fluid cylinder with equation of state given by Eq.
(3) with $p_1=0$. The system of differential equations following from the
Papapetrou form of the metric is simpler than the system corresponding to
the Lewis form. The general stationary solution is similar to the Davidson
solutions and to the general static solution. It is easier to challenge
rotation than the case $p_1\neq 0$ in the equation of state. The solution is
algebraic except at $\gamma =3/2$ where it becomes exponential. It has two
free parameters $p_0$ and $a_0$ describing the central pressure and
rotational effects. The Davidson one-parameter solutions are obtained when
Eq. (22) is imposed on the parameters.

The crucial technical point is the choice of $H$, valid only in the
stationary case. This is a partial fixing of the gauge, because an arbitrary
function still remains in the solution. It is advantageous to shift the
gauge function from $H$ to $A$ and $F$. The two realistic regions of $\gamma 
$ are $\left( 1.41,1.5\right) $ and $[1.5,2]$. A regular solution requires
different choices of $F$ in them. Therefore, the gauge should be fixed
completely as a last step. As a whole, our results confirm the qualitative
picture drawn in Refs. \cite{twentyfive,twentysix,twentyseven,twentyeight}.
It is strange that realistic solutions exist only for highly relativistic
perfect fluid and there is a gap between it and dust.

We have shown that the cosmological and the G\"odel solutions follow
directly from the general solution. The dust solution is easily restored
from the original system of equations. The same is true for dust on
cosmological background. We have tried to solve the controversy between a
recent paper \cite{thirtythree} and the older research on this topic.

Interesting features of the general solution are the following. The
magnitude of the vorticity vector is a square root of the pressure. As a
consequence, the angular velocity depends only on the central pressure and
not on $a_0$. The curvature invariants are well-behaved because the most
dangerous terms are proportional to powers of the acceleration vector's
magnitude. The latter decrease monotonically and vanishes at infinity. CTC
appear for any $\gamma $ inside the realistic region. The parameter $a_0$
commands the distance at which $g_{\varphi \varphi }$ becomes negative. More
vigorous rotation invokes CTC nearer to the axis.

\end{document}